\documentclass[fleqn,usenatbib]{mnras}
\usepackage{amsmath}
\usepackage{graphicx}
\usepackage{amssymb,txfonts}

\newcommand{\msun}{{\rm M}_{\sun}}

\newcommand{\g}{$\gamma$}

\newbox\grsign \setbox\grsign=\hbox{$>$} \newdimen\grdimen \grdimen=\ht\grsign
\newbox\simpropbox
\setbox\simpropbox=\hbox{\raise.5ex\hbox{$\propto$}\llap
   {\lower.5ex\hbox{$\sim$}}}\ht2=\grdimen\dp2=0pt
\def\simprop{\mathrel{\copy\simpropbox}}

\title[Variable jet Lorentz factors]{Variable jet Lorentz factors can explain soft self-absorbed radio spectra of accreting black-holes}

\author[A.~A. Zdziarski]
{Andrzej A. Zdziarski\thanks{E-mail: aaz@camk.edu.pl}\\
Nicolaus Copernicus Astronomical Center, Polish Academy of Sciences, Bartycka 18, PL-00-716 Warszawa, Poland\\}

\date{Accepted 2019 August 12. Received 2019 August 12; in original form 2019 July 12.}

\pagerange{\pageref{firstpage}--\pageref{lastpage}}
\pubyear{2019}

\begin{document}

\maketitle

\label{firstpage}

\begin{abstract}
We study the effect of variable jet bulk Lorentz factors, i.e., either jet acceleration or deceleration, on partially synchrotron self-absorbed radio spectra from cores of radio-loud active galactic nuclei and black-hole binaries in the hard state. In about a half of quasars and radio galaxies, their core radio spectra are observed to be soft, i.e., have the spectral index of $\alpha<0$. If they are emitted by jets with constant Lorentz factors, that softness implies deposition of large amounts of energy at large distances from the centre. We show here that such soft spectra can be explained without that energetic requirement by emission of jets with the Doppler factor increasing with the distance. This can happen for either jet acceleration or deceleration, depending on the jet viewing angle. We find our model can explain the quiescent radio to X-ray spectra of the BL Lac objects Mrk 421 and Mrk 501.
\end{abstract}

\section{Introduction}
\label{intro}

The model of \citet{bk79} was developed in order to account for flat radio spectra of jets, $F(\nu) \simprop \nu^\alpha$ with $\alpha\sim 0$ [where $F(\nu)$ is the energy flux per unit photon frequency], which spectra are common in the cores of quasars, BL Lac objects and radio galaxies (see, e.g., the compilation of \citealt{yuan18}), as well as in the hard state of BH binaries (e.g., \citealt{fender00}). Such spectra most likely result from a superposition of synchrotron emission that is self-absorbed up to some distance, $z\simprop \nu^{-1}$, along the jet and it becomes optically thin emission above it. Therefore, most of the emission at a given frequency $\nu$ is radiated from that distance. As the flat radio spectra can extend down to $\sim$100 MHz (e.g., \citealt{zywucka14}), this implies the partially self-absorbed emission up to rather large distances. 

In steady state, the spectral index of partially synchrotron self-absorbed emission depends on the spatial distributions of both the relativistic electrons and magnetic fields. If these distributions in the comoving frame are given by power laws of the distance along a conical jet with constant velocity, with the electron density of $N\propto z^{-a}$, and the magnetic field strength of $B\propto z^{-b}$, the spectral index is a function of the two spatial indices
\begin{equation}
\alpha=\frac{5 a+3 b+2(b-1)p-13}{2a-2+b(p+2)}
\label{alpha}
\end{equation}
(\citealt{konigl81}; the explicit expression is given in \citealt*{z19}). For a conserved electron number at a given energy and conserved magnetic energy flux in a conical jet with a constant speed, the two corresponding indices are $a=2$ and $b=1$, respectively, corresponding to $\alpha=0$. If the electrons suffer radiative and adiabatic losses, $a>2$, and if the magnetic energy is dissipated, $b>1$. In either case, the spectrum becomes harder, $\alpha>0$, corresponding to the emission weakening with the distance (with respect to the conserved case). Thus, spectra harder than $\alpha=0$ appear naturally in the presence of dissipation of the jet internal energy. On the other hand, spectra softer than $\alpha=0$ require the energy fluxes of either electrons or magnetic field to increase along the jet. The latter would require an external source of magnetic flux, which appears unlikely. Thus, spectra with $\alpha<0$ require deposition of substantial amounts of energy in relativistic electrons at large distances. A possible explanation of that is advection of the relativistic electrons from downstream in the jet, as proposed by \citet{z19}. That, however, requires the absence of effective adiabatic losses, which can happen if the energy lost in that process is used for reacceleration of the relativistic electrons present in the jet \citep{z19}. We are not certain how common that phenomenon is in astrophysical jets.

Soft spectra with $\alpha<0$ are found to be emitted by radio cores of about a half of the large sample (752 radio galaxies and 455 radio quasars) of radio-loud AGN analysed by \citet{yuan18}. Namely, they found the mean of $\langle \alpha \rangle=-0.001$ and the large standard deviation of $\sigma_\alpha=0.397$ for the core spectra of the combined sample. Two examples of blazars with soft radio spectra are the BL Lacs Mrk 501 and Mrk 421, which have $\alpha\approx -0.2$ in their quiescent states \citep{abdo11a,abdo11b}.

Radio spectra with $\alpha$ significantly lower than 0 could, in principle, originate from optically-thin synchrotron emission by relativistic electrons with hard distributions, with $p=1-2\alpha$, where $p$ is the index of the electron power-law distribution ($p<1$ would violate the particle number conservation). Given the relative smoothness and symmetry of the distribution of the indices \citep{yuan18}, the origin of those spectra from partially self-absorbed emission is very likely. Furthermore, the widely observed phenomenon of the core shift \citep{pushkarev12} is well explained by the synchrotron self-absorption and would not occur for purely optically thin emission. We also note that the spectra observed from outside the cores, which almost certainly originate from optically-thin synchrotron, are much softer, with the total spectra distributed as $\langle\alpha\rangle\approx -0.79$, $\sigma_\alpha\approx 0.25$ \citep{yuan18}.

Most of previous calculations of jet emission assumed the jet to have a constant bulk velocity. However, astrophysical jets can be both accelerated and decelerated. Possible acceleration mechanisms include conversion of a fraction the Poynting flux (e.g., \citealt{komissarov11}) or a fraction of the adiabatic losses of relativistic particles \citep{laing04} into bulk motion. Deceleration can occur due to interaction with the surrounding media. Observationally, jet acceleration has been measured in blazars, see, e.g., \citet{lister19}. 

In this Letter, we study the effect of a variable jet bulk Lorentz factor on partially self-absorbed synchrotron spectra. We find that indeed either jet acceleration or deceleration can lead to soft spectra with $\alpha<0$.


\section{The jet model}
\label{jet}

We follow here the general approach of \citet{bk79}, for which detailed formalism was developed in \citet*{zls12} and \citet{z19}. We define a dimensionless distance along the jet, $\xi\equiv z/z_0$, where $z_0$ is the location of the onset of the emission. We use a mono-energetic approximation to the synchrotron emission and absorption and define a dimensionless photon energy in the jet frame, $\epsilon$,
\begin{equation}
\epsilon= \frac{B}{B_{\rm cr}}\gamma^2,\quad 
\epsilon(E,\xi)\equiv \frac{E(1+z_{\rm r})}{\delta(\xi) m_{\rm e}c^2},\quad \delta(\xi)\equiv \frac{1}{\Gamma(\xi)[1-\beta_{\rm j}(\xi)\cos i]},
\label{epsilon}
\end{equation}
where $\gamma$ is the electron Lorentz factor, $\delta$ is the Doppler factor, $\beta_{\rm j}c$ is the jet velocity, $z_{\rm r}$ is the redshift, $B$ and $\gamma$ are measured in the jet frame, $B_{\rm cr}$ is the critical magnetic field, $E$ is the observed dimensional photon energy, and $m_{\rm e}c^2$ is the electron rest energy. We allow the jet to have a variable bulk Lorentz factor, $\Gamma(\xi)$, and an arbitrary distance dependence of the jet radius, $r(\xi)$. 

The resulting form of the synchrotron spectrum valid across the partially self-absorbed to optically-thin regimes is given by equations A4 and A5 of \citet{z19},
\begin{align}
&F_{\rm S}(E)=\left(m_{\rm e} c\over h\right)^3 {\upi (1+z_{\rm r})c C_1 z_0 (\epsilon\delta)^{5/2}\sin i\over 6 C_2 D_L^2} \times\nonumber\\
&\qquad \int_{\xi_0}^{\xi_{\rm max}} {\rm d}\xi\,r(\xi)\left[\delta(\xi)B_{\rm cr}\over B(\xi)\right]^{1/2} \left\{1-\exp\left[-\tau_{\rm sa}(E,\xi)\right]\right\},
\label{1D}\\
&\tau_{\rm sa}(E,\xi)= \frac{C_2 \upi \sigma_{\rm T} r(\xi)B(\xi)}{\alpha_{\rm f} B_{\rm cr}\epsilon(E,\xi)^2 \delta(\xi) \sin i} N\left(\sqrt{\epsilon(E,\xi) B_{\rm cr}\over B(\xi)},\xi\right),
\label{tausyn}
\end{align}
where $i$ is the jet viewing angle, $N(\gamma,\xi)$ is the steady-state electron distribution in the comoving frame, which is now assumed to be a power law in $\gamma$ with the index $p$, $N\propto\gamma^{-p}$, $C_1(p)\sim 1$ and $C_2(p)\sim 1$ are normalization factors for the synchrotron emissivity and absorption coefficient, respectively (weakly dependent on $p$, see, e.g., \citealt{zls12}), $\tau_{\rm sa}$ is the self-absorption optical depth of the jet viewed at the angle $i$, the product $\epsilon\delta$ equals the dimensionless photon energy in the BH frame (and thus it is independent of $\xi$; see equation \ref{epsilon}), $h$ is the Planck constant, $\sigma_{\rm T}$ is the Thomson cross section, $\alpha_{\rm f}$ is the fine-structure constant, $D_L$ is the luminosity distance, and $\xi_0(\epsilon)$ is the minimum $\xi$ at which a given observed photon energy, $E$, can be produced by the synchrotron process in a given magnetic field, given by the maximum of 1 and the solution of
\begin{equation}
\frac{\epsilon(E,\xi_0) B_{\rm cr}}{B(\xi_0)}=\gamma_0^2,
\label{ximin}
\end{equation}
where $\gamma_0$ is the minimum Lorentz factor in the distribution of the relativistic electrons. The normalization of equation (\ref{1D}) assumes that the energy unit in the energy flux per unit photon energy, ${\rm d}F_{\rm S}/{\rm d}E$, is the same as the unit of the photon energy, $E$. If the energy units of photon energy in $F_{\rm S}$ were different from that of $E$ then $F_{\rm S}$ should be multiplied by the ratio of the two units.

We assume electrons are accelerated by a magnetic process, with the rate $\dot \gamma_{\rm acc} m_{\rm e} c=\eta_{\rm acc} e B$, where $e$ is the electron charge. The acceleration is balanced by synchrotron losses with the efficiency factor $\eta_{\rm acc}$, which is usually $\la 1/2\upi$. This yields
\begin{equation}
\gamma_{\rm max}(\xi) \approx \left[9 \eta_{\rm acc}B_{\rm cr} \over 4\alpha_{\rm f} B(\xi)\right]^{1/2}.
\label{gemax}
\end{equation}
The corresponding maximum jet-frame photon energy, $\epsilon_{\rm max}= \gamma_{\rm max}(\xi)^2 B(\xi)/B_{\rm cr}=9 \eta_{\rm acc}/(4\alpha_{\rm f})$, is independent of $B$ \citep*{gfr83}. We then assume that the local spectrum emitted at a $\xi$ is e-folded with that $\epsilon_{\rm max}$, i.e., we multiply the integrand in equation (\ref{1D}) by $\exp[-\epsilon(\xi) /\epsilon_{\rm max}]$.

\begin{figure}
\centerline{\includegraphics[width=\columnwidth]{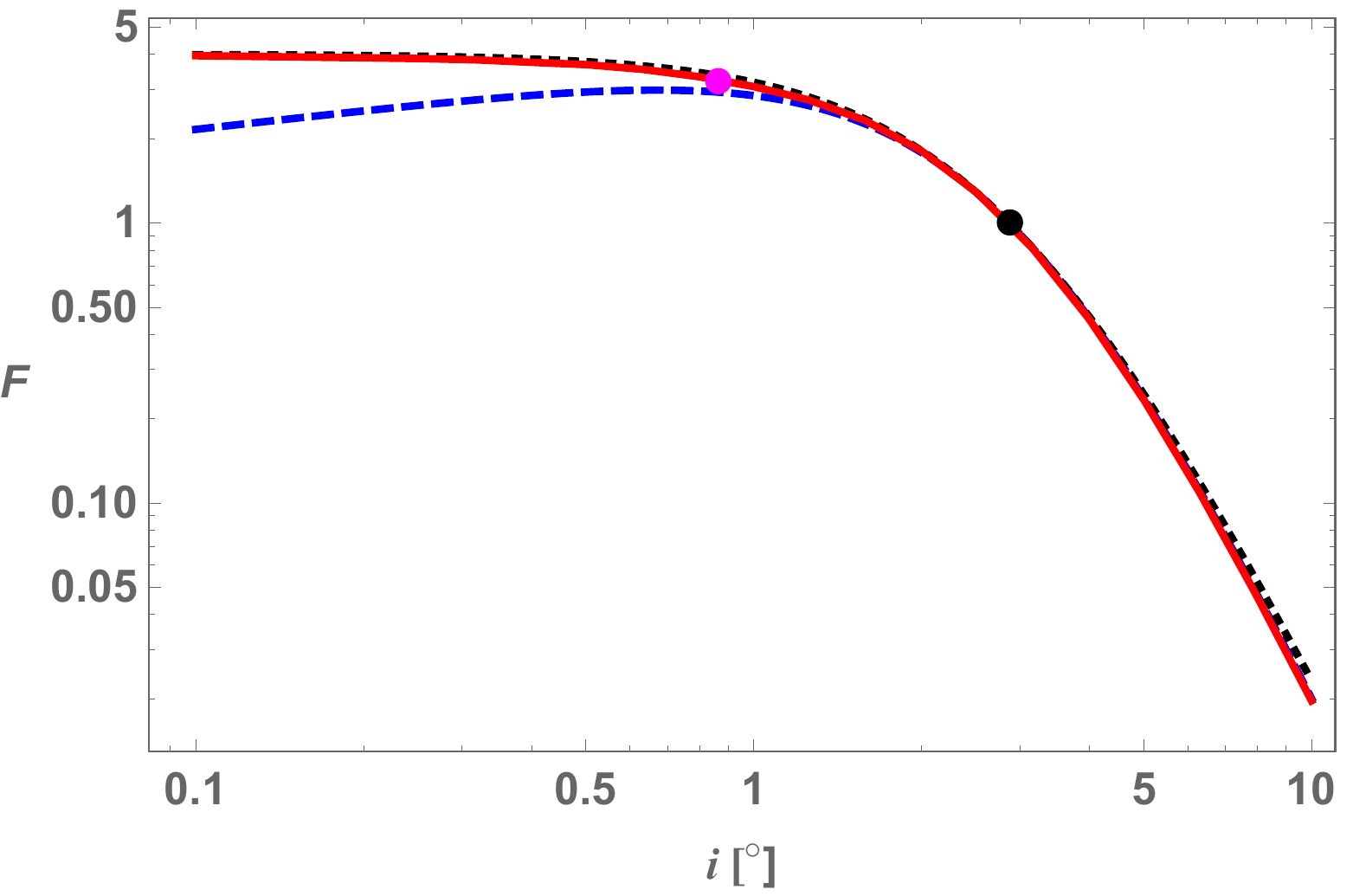}}
\caption{A comparison of the jet angular distributions of partially self-absorbed synchrotron emission of a conical jet with constant $\Gamma=20$, the half-opening angle of $\tan\Theta= 0.3\arcsin(1/\Gamma)$ (shown by the magenta dot) and $p=2.5$. The red solid curve shows the exact dependence for a partially optically-thick jet \citep{z16}. The blue dashed curve shows the approximation of equation (\ref{1D}), which becomes inaccurate only at small angles. The black dotted curve shows the distribution for an optically thin steady-state jet, i.e., $F\propto \delta_{\rm j}^2$, which is very close to the exact dependence. The flux is normalized to unity at $i= \arcsin(1/\Gamma)$ (shown by the black point), where $\delta=\Gamma$ and the angular dependence factor of the dashed curve, $\delta^{(7+3 p)/(4+p)}(\sin i)^{(p-1)/(4+p)}$, becomes equal to $\delta^2$.
}
\label{angular}
\end{figure}

\begin{figure}
\centerline{\includegraphics[width=\columnwidth]{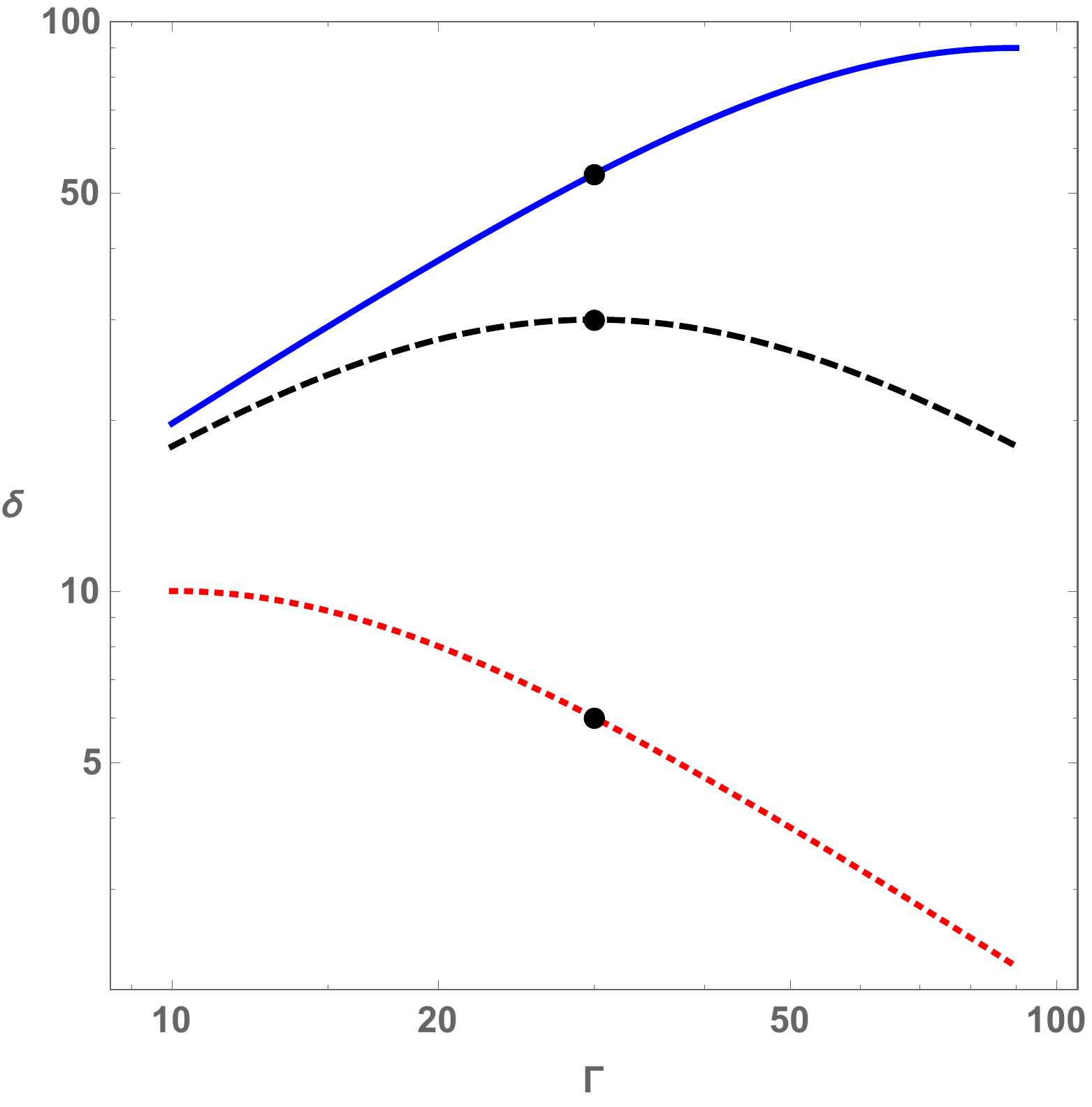}}
\caption{Examples of the $\delta(\Gamma)$ dependence for the initial $\Gamma_0=30$ (shown by the black points) and $\Gamma$ increasing and decreasing from $\Gamma_0$ by a factor of 3, for $i=\arcsin(1/\Gamma_0)$ (black dashed curve), $i=(1/3)\arcsin(1/\Gamma_0)$ (blue solid curve) and $i=3\arcsin(1/\Gamma_0)$ (red dotted curve). We see that for the canonical choice of the viewing angle, $i=\arcsin(1/\Gamma_0)$, either increasing or decreasing $\Gamma$ leads to a lowering the Doppler factor. Then either jet acceleration or deceleration lead to a hardening of the partially self-absorbed spectra, $\alpha>0$. On the other hand, a lower or higher viewing angle leads to $\delta$ increasing with increasing and decreasing $\Gamma$, respectively. Both cases can then result in soft partially self-absorbed spectra, with $\alpha<0$.
}
\label{dg1}
\end{figure}

The above formalism is based on integrating the radiative transfer equation (with a source function corresponding to synchrotron-emitting non-thermal electrons) over the projection on the plane of the sky of a conical jet viewed sideways in the observer's frame. Since we are concerned here with the viewing angles close to jet opening angles, we consider here the validity of the approach based on equations (\ref{1D}--\ref{tausyn}) for such angles. The flux dependence on the angle of partially self-absorbed synchrotron emission in the case of a conical jet with a half-opening angle $\Theta$, a constant $\Gamma$, and conserved electron distribution and magnetic energy flux (as in \citealt{bk79}) is $\propto \delta^{(7+3 p)/(4+p)}(\sin i)^{(p-1)/(4+p)}$ (see also \citealt{cawthorne91}). This implies that the jet emission is null at $i=0$ and has the maximum at an angle of a fraction of order unity of $i=\arcsin(1/\Gamma)$ \citep{cawthorne91}. However, this formalism ignores the emission of the head of the jet. That emission was analysed in \citet{z16}, who derived formulae allowing to calculate exact emission of a conical jet with the half-opening angle of $\Theta$ including the emission at angles lower than the jet half-opening angle, and obtained a simple analytical formula for the emission at $i=0$. Here, we show an example of the exact dependence a conical jet with $\tan\Theta= 0.3\arcsin(1/\Gamma)$ and $\Gamma=20$, see the red solid curve in Fig.\ \ref{angular}. We see that the maximum (rather than the minimum) of the emission is at $i=0$. The blue dashed curve shows the approximation of equations (\ref{1D}--\ref{tausyn}).  We see that the model of a conical jet viewed sideways still provides a good approximation to the exact dependence for viewing angles $i\gtrsim 0.3\arcsin(1/\Gamma)$, which justifies our use of those equations in the analysis below. (A similar comparison was shown in \citealt{z16} for $\Gamma=5/3$.)

We also show the exact dependence of the partially self-absorbed synchrotron emission is rather close to $\propto \delta^2$ (see the black dotted curve), i.e., that of an optically-thin steady state jet, which emission is $\propto \delta^{2-\alpha}$ \citep{bk79,lb85,sikora97}, where in the considered case $\alpha=0$. We note, however, that the flux dependence for $i\la \Theta$ of the exact formula depends on the value of $\Theta$, unlike that of $\delta^2$, and the agreement is best\footnote{The exact value of the flux from a jet with constant $\Gamma$ at $i=0$ is given by equation 14 of \citet{z16}. The $\delta^2$ approximation reaches the exact value at $i=0$ for the jet opening angles given by $\tan\Theta\approx 0.6 \sin(1/\Gamma)/(1+\beta)$ almost independently of $p$ (the exact formula can be calculated analytically).} for $\Theta\simeq 0.3/\Gamma$.

Given that the treatment based on equations (\ref{1D}--\ref{tausyn}) remains quite accurate down to a small fraction of the characteristic viewing angle of $i\approx 1/\Gamma$, we can proceed further and use the above formalism for the cases of accelerated and decelerated jets. However, since we will assume a variable $\Gamma$, we cannot use the $\delta^2$ approximation and have to instead use equations (\ref{1D}--\ref{tausyn}). Our goal here is to account for soft partially self-absorbed spectra, i.e., with $\alpha<0$. This requires to enhance (with respect to the constant $\Gamma$ case) the emission at large distances. Since $F\simprop \delta^2$, we need to consider jets with $\delta$ increasing with the distance from the jet origin. Thus, we show some examples of dependencies of $\delta(\Gamma)$ for the viewing angles close to $i=\arcsin(1/\Gamma)$. The solid curve in Fig.\ \ref{dg1} shows the case of the initial $\Gamma_0=30$ and $i=\arcsin(1/\Gamma_0)$ and $\Gamma$ increasing and decreasing from it. We see that either results in decreasing $\delta$. This means that the spectra from such a jet will become harder, $\alpha>0$, rather than softer. On the other hand, choosing a lower value of $i$ can lead to $\delta$ increasing with increasing $\Gamma$, and choosing a higher value of $i$ can lead to $\delta$ increasing with decreasing $\Gamma$, as shown in Fig.\ \ref{dg1} by the dashed and dotted curves, respectively. Since the usual estimate of $i=\arcsin(1/\Gamma_0)$ is based on a statistical argument, as well as accelerated or decelerated jets do not have a unique Doppler factor, the last two cases can correspond to sources with $\alpha<0$.

We intend here to show the effect of variable Lorentz factor on the partially self-absorbed synchrotron spectra. Therefore, we adopt simplest possible assumptions about the distribution of relativistic electrons, the spatial dependence of the magnetic field, and the shape of the jet. Namely, we still assume the standard dependencies of \citet{bk79},
\begin{equation}
N(\gamma,\xi)=N_0 \xi^{-2} \gamma^{-p},\quad B(\xi)=B_0 \xi^{-1},\quad r(\xi)=z_0 \xi \tan\Theta,
\label{NB}
\end{equation}
which lead to $\alpha=0$ for a constant $\Gamma$. While these dependencies neglect the effect of variable $\Gamma$ on the quantities in the comoving frame, and the possible conversion of the Poynting flux into the electron acceleration, they can serve to illustrate that effect on the self-absorbed spectra by perturbing only one parameter, $\Gamma$. For simplicity, we also assume the jet to be conical. (We will consider self-consistent continuous jet models in our work in preparation, where we also will include effects of radiative losses and Compton scattering.) In order to parametrize the Lorentz factor changing along the jet, we assume a power-law dependence and parametrize $\Gamma$ in terms of its final value, $\Gamma_{\rm fin}$, as 
\begin{equation}
\Gamma=\Gamma_{\rm fin} (\xi/\xi_{\rm max})^q,
\label{Gamma}
\end{equation}
where $q>0$ ($<0$) corresponds to accelerated (decelerated) jets. 

\section{Comparison with data}
\label{data}

We compare our model synchrotron spectrum with those of some quiescent spectra of blazars, namely Mrk 421 and Mrk 501, which are high-synchrotron peaked BL Lac object at the redshifts of $z=0.0308$ and 0.03364, respectively. We have chosen those objects because they have soft radio-to-IR spectra with $\alpha\approx -0.2$, as well as they are ones of the brightest sources in their class, their continua from radio to hard X-rays are well covered and of high quality, see \citet{abdo11b} and \citet{abdo11a}, respectively. The data for Mrk 421 have been obtained and averaged over a single observational campaign lasting about 1.5 yr, during which the measured fluxes were almost constant \citep{abdo11b}. The observational campaign for Mrk 501 lasted 4.5 months \citep{abdo11a}. We use the luminosity distances of $D_L=133$ Mpc and 146 Mpc, respectively, corresponding to $H_0=71$ km s$^{-1}$ Mpc$^{-1}$ and $\Omega_\Lambda =0.73$. The BH mass of Mrk 421 was measured via the $M$-$\sigma$ relation by \citet*{barth03} as $\log_{10} (M/\msun)=8.28\pm 0.11$; here we assume $M=2\times 10^8\msun$, corresponding to $r_{\rm g}\approx 3\times 10^{13}$ cm. The BH mass of Mrk 501 was estimated by the same method by \citet*{barth02} as $M\approx (0.9$--$3.5)\times 10^9\msun$. Here, we assume $M=10^9\msun$, though we note that the actual mass may be lower \citep{rieger03}.

\begin{figure}
\centerline{\includegraphics[width=\columnwidth]{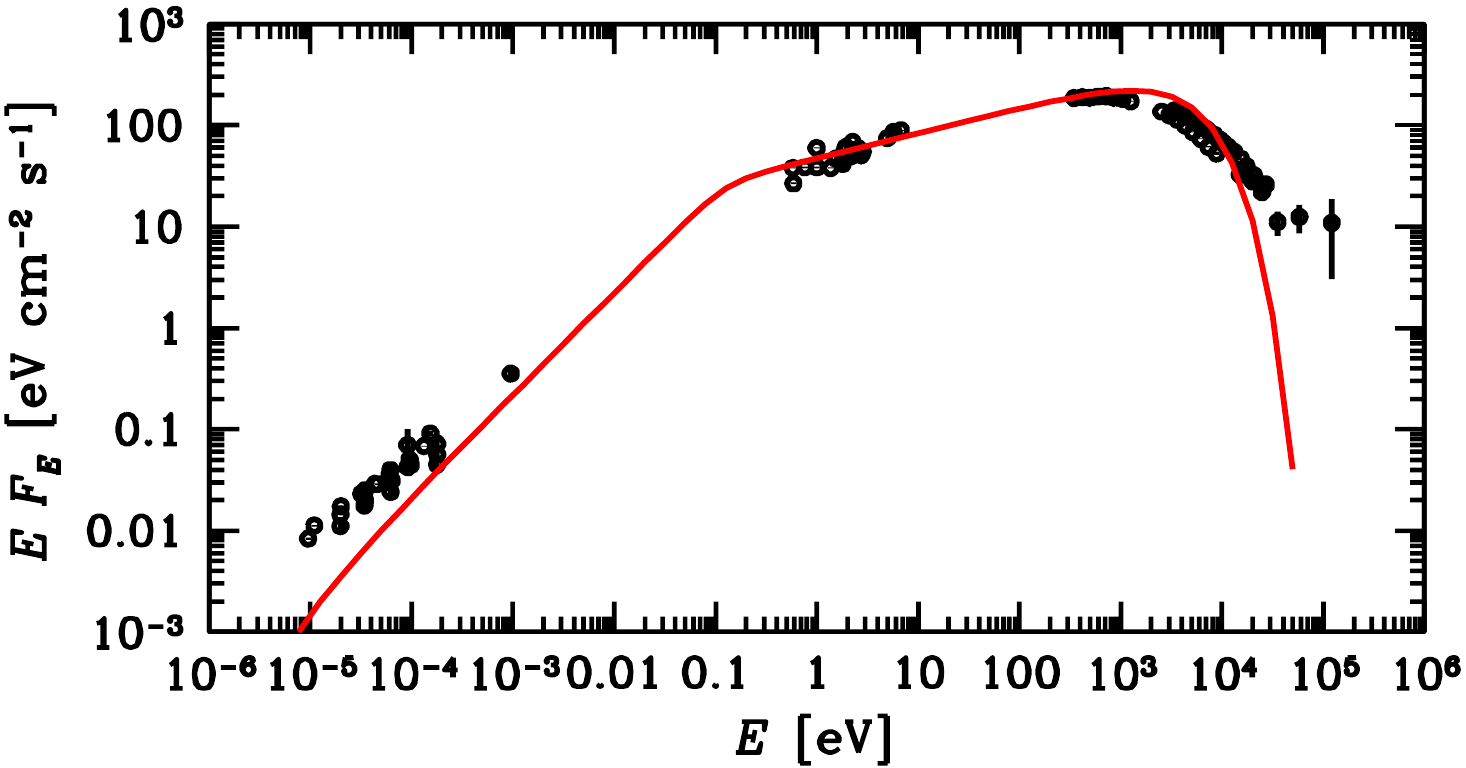}}
\caption{The radio-to-X-ray spectrum of the quiescent state of Mrk 421 \citep{abdo11b} compared to the synchrotron model with constant $\Gamma$, and in which the electron distribution is maintained along a conical jet with constant velocity and conserved energy flux of toroidal magnetic field. We see that this model fails to reproduce the radio--mm part of the spectrum. The model parameters are $\Gamma=30$, $\Theta=0.2/\Gamma$, $z_0=20 r_{\rm g}$, $z_{\rm max}=10$ pc, $B_0=40$ G, $N_0=3\times 10^9$ cm$^{-3}$, $p=2.5$, $i=0.5/\Gamma$, $\eta_{\rm acc}=7\times 10^{-7}$.
}
\label{f:bk79}
\end{figure}

\begin{figure}
\centerline{\includegraphics[width=\columnwidth]{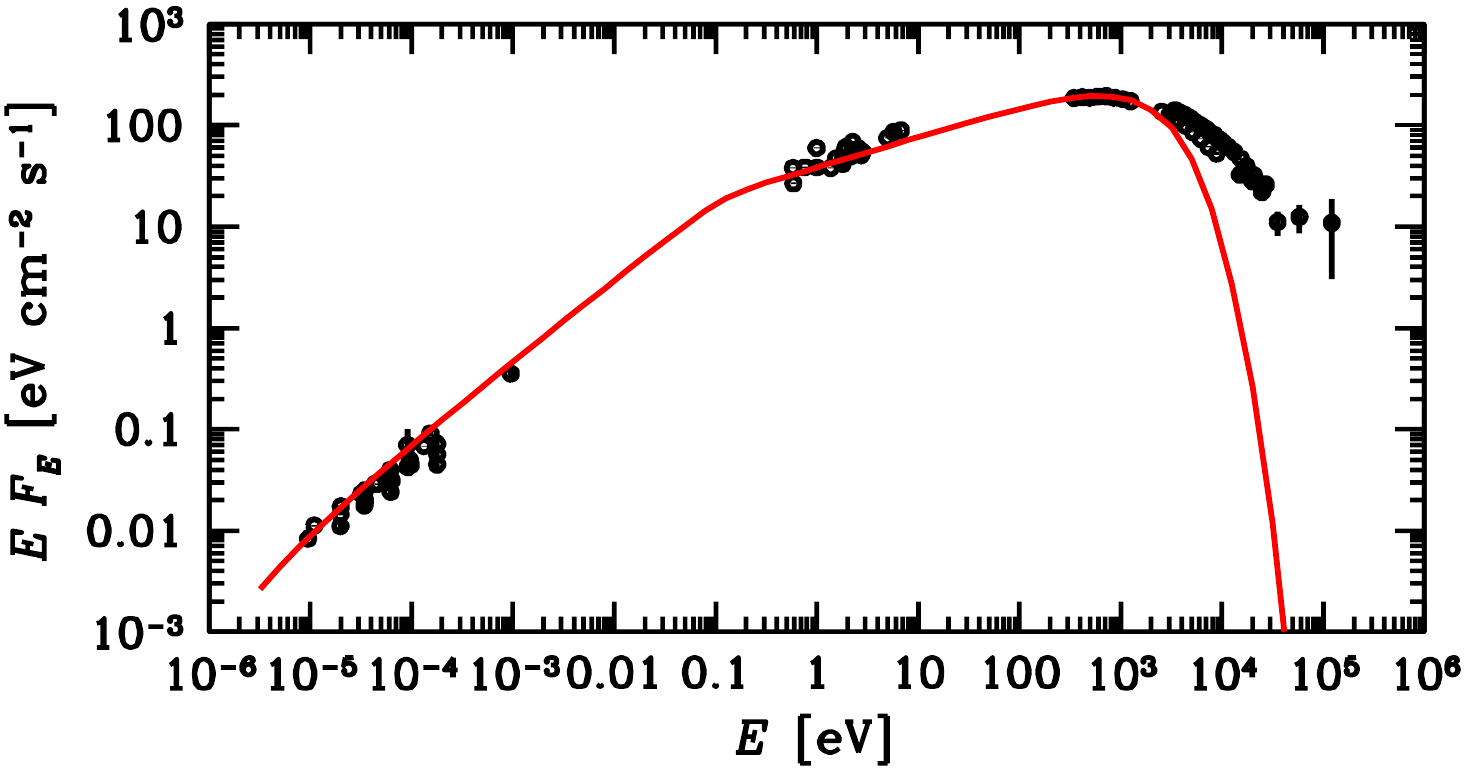}}
\caption{The radio-to-X-ray spectrum of the quiescent state of Mrk 421 compared to the synchrotron model with an accelerated $\Gamma$ with $q=0.1$ up to the terminal value of $\Gamma_{\rm fin}=30$ (which corresponds to the initial $\Gamma\approx 8$), and $i=0.5/\Gamma_{\rm fin}$, $\Theta=0.2/\Gamma_{\rm fin}$. The remaining parameters are $p=2.4$, $N_0=1.7\times 10^{10}$ cm$^{-3}$, $z_0=20 r_{\rm g}$, $z_{\rm max}=100$ pc, $B_0=40$ G, $\eta_{\rm acc}=7\times 10^{-7}$. 
}
\label{spectrum421}
\end{figure}

\begin{figure}
\centerline{\includegraphics[width=\columnwidth]{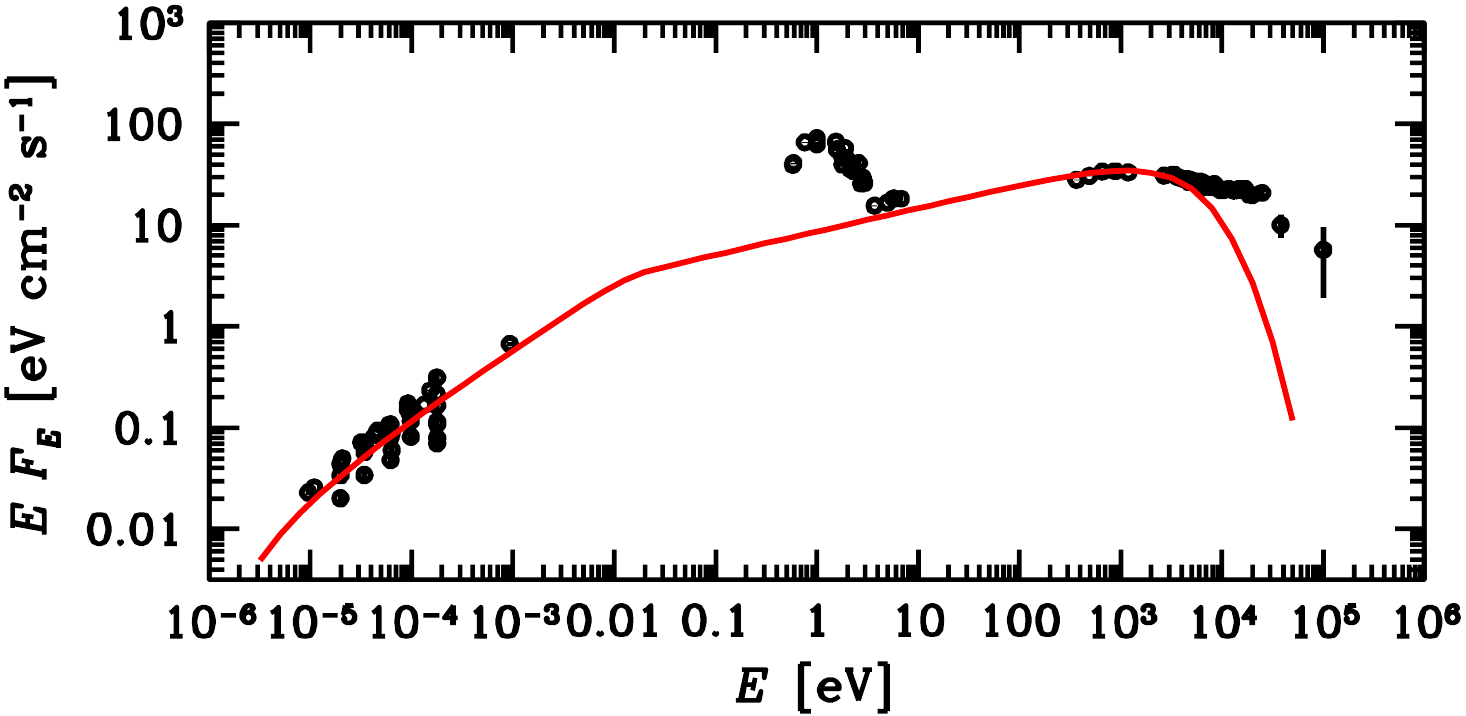}}
\caption{The radio-to-X-ray spectrum of the quiescent state of Mrk 501 compared to the synchrotron model with a decelerated $\Gamma$ with $q=-0.15$ from $\Gamma=48$ at $z_0=60 r_{\rm g}$ to the terminal value of $\Gamma_{\rm fin}=10$, and $i=3/\Gamma_{\rm fin}$, $\Theta=0.1/\Gamma_{\rm fin}$. The remaining parameters are $p=2.54$, $N_0=3.5\times 10^{8}$ cm$^{-3}$, $z_{\rm max}=100$ pc, $B_0=400$ G, $\eta_{\rm acc}=5\times 10^{-5}$. The feature around 1 eV is the emission of the host galaxy, not subtracted from the average spectrum \citep{abdo11a}.
}
\label{spectrum501}
\end{figure}

Fig.\ \ref{f:bk79} shows an example of the fit of the model to the spectrum of Mrk 421 assuming $q=0$, i.e., a constant $\Gamma$. We assume $\Theta=0.2/\Gamma$ (following \citealt{pushkarev09,clausen13}). Since this model yields $\alpha=0$ in the partially self-absorbed part, it cannot reproduce the radio-to-IR part of the spectrum. On the other hand, we find we can indeed reproduce the spectrum of Mrk 421 with an accelerated jet, with $q=0.1$, as shown in Fig.\ \ref{spectrum421}, which caption also gives all the parameters assumed in the model.

As discussed above, we should also be able to obtain soft radio-to-IR self-absorbed spectra with a decelerated jet. We indeed find a good model of the spectrum of Mrk 501 for $q=-0.15$, as shown in Fig.\ \ref{spectrum501}, which caption also gives all the parameters assumed in the model.

While we have fitted Mrk 421 with an accelerating jet, \citet{potter13b} (based on the model of \citealt{potter13a}) found they can model it by a significant deceleration. We stress, however, that our fits were intended to illustrate the possibility of reproducing soft radio-to-IR spectra by changing bulk Lorentz factor and are neither self-consistent nor unique. We could have, in fact, fit Mrk 421 with a decelerating jet and Mrk 501 with an accelerating one. Furthermore, we note that \citet{potter13b} managed to obtain a good fit to the soft radio spectrum of Mrk 421 by assuming it was optically thin synchrotron emission. For that emission, changing the bulk Lorentz factor does not change the slope of the observed spectrum (unlike the case of partially self-absorbed emission) and thus the deceleration used in that model does not affect the value of $\alpha$. As we argued in Section \ref{intro}, while reproducing spectra with $\alpha<0$ by optically thin emission may work in some individual cases, it cannot account for most of the observational radio core spectra of the sample of \citet{yuan18} with $\langle\alpha\rangle\approx 0.0$, $\sigma_\alpha\approx 0.4$.

Furthermore, as we stated in Section \ref{jet}, equation (\ref{NB}), which parametrizes the electron density, magnetic field strength and the jet radius, is purely phenomenological. Our simple model also ignored the \g-ray part of the spectra, whose form imposes constraints on the models via the requirement that the synchrotron self-Compton emission does not exceed the observed spectrum. Reproducing the broad-band spectrum from radio to $\gamma$-rays requires significantly more complex models, which is beyond the scope of our Letter.

\section{Conclusions}
\label{conclusions}

We have considered partially synchrotron self-absorbed emission of jets with a variable Lorentz factor. Our goal has been to explain the origin of soft flat radio spectra, with $\alpha<0$. We have shown such spectra can be obtained if the Doppler factor of an extended jet increases with the increasing distance from the jet origin. An increasing Doppler factor can be achieved for either accelerated and decelerated jets, depending on the initial viewing angle.

We have adopted a formalism based on the integration of the non-thermal source function along the projected area of the jet, assumed to be conical, and made simple assumptions about the power-law dependencies of the electron density and the magnetic field strength. 

We have applied our model to two BL Lac objects, Mrk 421 and Mrk 501. We have shown that their broad-band radio to X-ray spectra can be well fitted by the model, assuming relatively slow changes of the bulk Lorentz factor of their jets. These models are clearly not unique, and are intended to show the viability of the general model.

\section*{ACKNOWLEDGEMENTS}

I thank Marek Sikora for valuable discussions and suggestions, {\L}ukasz Stawarz for providing the quiescent spectra of Mrk 421 and Mrk 501, and the referee for valuable comments. This research has been supported in part by the Polish National Science Centre grant 2015/18/A/ST9/00746.

\label{lastpage}

\end{document}